\documentclass{aa}  
\usepackage{graphicx}
\usepackage{txfonts}
\usepackage{natbib} 
\usepackage{float}
\usepackage{rotating}
\usepackage{amsmath}
\bibpunct{(}{)}{;}{a}{}{,} 
\begin{document}
   \title{Protoplanetary disk lifetimes vs stellar mass \\ and possible implications for giant planet populations}
   \author{\'Alvaro Ribas\inst{\ref{esac},\ref{cab},\ref{insa}}
     \and
     Herv\'e~Bouy\inst{\ref{cab}} 
     \and
     Bruno~Mer\'{i}n\inst{\ref{esac},\ref{herschel}}
     }

   \offprints{\'A. Ribas}
	\institute{
	European Space Astronomy Centre (ESA), P.O. Box, 78, 28691 Villanueva de la Ca\~{n}ada, Madrid, Spain\label{esac} \\
              \email{aribas@cab.inta-csic.es}
         \and
        Centro de Astrobiolog\'{i}a, INTA-CSIC, P.O. Box - Apdo. de correos 78, Villanueva de la Ca\~nada Madrid 28691, Spain\label{cab} 
        \and
	Ingenier\'ia y Servicios Aeroespaciales-ESAC, P.O. Box, 78, 28691 Villanueva de la Ca\~{n}ada, Madrid, Spain\label{insa}
        \and
	Herschel Science Centre, ESAC-ESA, P.O. Box, 78, 28691 Villanueva de la Ca\~{n}ada, Madrid, Spain\label{herschel}}
   \date{Received 22 August 2014; Accepted 19 January 2015}

 
  \abstract
{}%
   {We study the dependence of protoplanetary disk evolution on stellar mass using a large sample of young stellar objects in nearby young star-forming regions.}
   {We update the protoplanetary disk fractions presented in our recent work (paper I of this series) derived for 22 nearby ($<$\,500\,pc)  associations between 1 and 100\,Myr. We use a subsample of 1\,428 spectroscopically confirmed members to study the impact of stellar mass on protoplanetary disk evolution. We divide this sample into two stellar mass bins (2\,M$_{\odot}$ boundary) and two age bins (3\,Myr boundary), and use infrared excesses over the photospheric emission to classify objects in three groups: protoplanetary disks, evolved disks, and diskless. The homogeneous analysis and bias corrections allow for a statistically significant inter-comparison of the obtained results.} 
   {We find robust statistical evidence of disk evolution dependence with stellar mass. Our results, combined with previous studies on disk evolution, confirm that protoplanetary disks evolve faster and/or earlier around high-mass ($>$\,2\,M$_{\odot}$) stars. We also find a roughly constant level of evolved disks throughout the whole age and stellar mass spectra. }
   {We conclude that protoplanetary disk evolution depends on stellar mass. Such a dependence could have important implications for gas giant planet formation and migration, and could contribute to explaining the apparent paucity of hot Jupiters around high-mass stars.}

   \keywords{Planetary systems: protoplanetary disks, planet-disk interactions -- stars: formation --  (stars:) planetary systems -- stars: pre-main-sequence}

   \maketitle


\section{Introduction}

Studying the evolution of protoplanetary disks around young stellar objects (YSOs) is crucial to better comprehend how planets form. Understanding the effects of fundamental parameters such as age, stellar mass, environment, or binarity provides insight into the planet formation processes, and could shed light onto important relations between planetary systems and their stellar hosts. Several studies on this topic \citep[see e.g.][]{Williams2011,Alexander2013,Espaillat2014} have already drawn a comprehensive picture of protoplanetary disk evolution. 

In the first paper of this series \citep[][hereafter R14]{Ribas2014} we presented a large sample of $\sim$\,2\,300 spectroscopically confirmed YSOs in 22 young ($<$ 100\,Myr) and nearby ($<$\,500\,pc) star-forming regions and associations; R14 used this large sample with very low contamination and high completeness levels to derive accurate disk fractions and protoplanetary disk dissipation timescales. Their results, in good agreement with previous estimates \citep{Haisch2001,Hernandez2007,Hernandez2008,Mamajek2009,Fedele2010,Murphy2013} served as a benchmark to test the reliability and robustness of the sample and methods used in R14.

The evolution of protoplanetary disks and planets are closely connected as the disk properties define the initial conditions of planet formation, and the planet evolution subsequently affects the disk properties \citep[e.g.][]{Kley2012,Baruteau2013}. The disk lifetime is a fundamental parameter to understand the interplay between a disk and its planets, since it determines the time available for the planets to form and migrate \citep[e.g.][]{Burkert2007,Alexander2009}. A dependence of the disk lifetime on the stellar mass could have strong implications for planet formation and migration theories \citep[see e.g.][for recent reviews on this topic]{Raymond2013,Helled2013,Baruteau2013}, and ultimately result in different planetary populations around low-mass and high-mass stars. Previous studies have already found hints of protoplanetary disks evolving faster around massive stars \citep[e.g.][]{Carpenter2006,Dahm2007,Kennedy2009,Roccatagliata2011,Fang2012, Yasui2014}, but these studies focused on individual regions and/or were potentially affected by different systematics (e.g., low-number statistics, sensitivity biases, contamination from background giant stars, different completeness levels). As a result,  a direct comparison of their results is difficult and a statistically robust confirmation of a possible dependence of the disk lifetime with stellar mass has been elusive so far.

In this second paper of the series, we study the influence of stellar mass on disk evolution using the large and homogeneous sample compiled in R14.  Section~\ref{sec:previous_sample} updates the study on R14. In Sect.~\ref{sec:masses_and_ages} we describe the methodology and criteria used to analyze the disk evolution dependence with stellar mass, and the obtained results are shown in Sect. \ref{sec:results}. The implications of this study are discussed in Sect.~\ref{sec:discussion}.

\section{Updates on paper I}\label{sec:previous_sample}

\subsection{Description of the sample and bias corrections}

We first make a short summary of the sample compilation and completeness corrections in R14. The dataset was obtained by combining several stellar population studies of different young regions and associations (see Table 1 in R14 for a complete list of references). These works used a variety of methods to confirm membership depending on stellar masses (e.g. presence of lithium in the spectra, accretion, or X-ray properties). Every object also has a spectroscopic measurement of its spectral type, minimizing the contamination from background giant stars and extragalactic sources. Moreover, these studies were specifically designed to make a complete census of the stellar content over a given mass range and should not be biased towards excess-bearing stars, for example. We therefore expect the compiled sample to be at least representative of (if not complete) the stellar population of each of the considered regions. The covered stellar mass spectrum ranges from O-type to late M-type stars in all regions.  We note that R14 did not re-estimate or discuss membership for any target.

Our main aim in R14 was to derive accurate disk fractions for the considered regions via infrared (IR) excesses. This dependence on the availability of IR data could bias the sample, and two additional steps were considered to prevent this:

\begin{itemize}

\item spatial correction, i. e., only sources within the field of view of IR surveys (Spitzer or WISE, see below) were considered. Given that disks are identified via IR measurements, sources lacking these data could be incorrectly classified as ``diskless'' (since no IR excess is found), even though there is no information to classify them (we do not know whether these sources have IR excess or not). 
We gathered IR data from the all-sky WISE survey \citep{WISE} to study associations from the SACY sample \citep{Torres2008}, and therefore they were not affected by this problem. For the rest of the regions, however, we used data from different \textit{Spitzer} Space Telescope programs and surveys, which were limited in their spatial coverage. In these cases, we checked the position of each source and did not considered those outside the respective \textit{Spitzer} map;

\item sensitivity correction, i. e. only sources with predicted photospheric level above the sensitivity limit of the corresponding observations were considered. Sources with disks have IR excesses, and therefore are easier to detect than naked stars (a process known as the Malmquist bias). We therefore considered only objects with predicted stellar emission (from photospheric models scaled to match the observed fluxes) above the corresponding sensitivity limit and wavelength. We also took into account the dependence of this limit with the corresponding photometric band and survey. For near-IR observations, most of the stars fulfill this requirement, and therefore the sample is expected to be complete down to mid or late M-type stars. In the mid-IR, the sensitivity of the observations is more restrictive and the completeness level drops down between late K- and mid M-type stars, depending on the distance of the region.

\end{itemize}

The result of this process is a representative sample of young (1-100\,Myr) nearby ($<$\,500\,pc) star-forming regions and associations with minimum biases.

\subsection{Updates in paper I}

\citet{Ribas2014} identified IR excesses via excess significance, defined as:

\begin{equation}\nonumber
  \chi_{\lambda} ={\rm \frac{F_{observed, \lambda } - F_{photosphere, \lambda }}{\sigma_{{\rm Observed},\lambda}}}
\end{equation}

where ${\rm F_{observed, \lambda }}$ is the observed flux, ${\rm F_{photosphere, \lambda }}$ is the corresponding model photospheric flux, and $\sigma_{{\rm Observed},\lambda}$ is the error at the corresponding wavelength ($\lambda$). In R14, bands with $\chi \ge 5$ were considered to harbor excesses. This procedure allowed us to compute disk fractions for each region at different wavelengths after correcting for sensitivity and spatial completeness (see R14 for a description of the method used).

\begin{figure}
  \caption{Ratio of the total uncertainty (observational and systematic) over the observed flux as a function of magnitude.  IRAC3, IRAC4, and MIPS1 bands are shown. Sources from discarded regions (IC348 and Ophiuchus) are represented as red crosses, the rest of the sources as black dots. The black solid line corresponds to the homogenized uncertainty levels of 7\,\% (all IRAC bands) and 12\,\% (MIPS1). The black dashed line shows the sensitivity limit for each band, as indicated by the observing programs. Objects falling beyond this limit in the MIPS1 plot are caused by noise (uncertainties in T$_{\rm eff}$, $A_{V}$,...) in the corresponding photosphere fit, and were kept in the sample.}
\label{fig:unc_dist}
\includegraphics[width=\hsize]{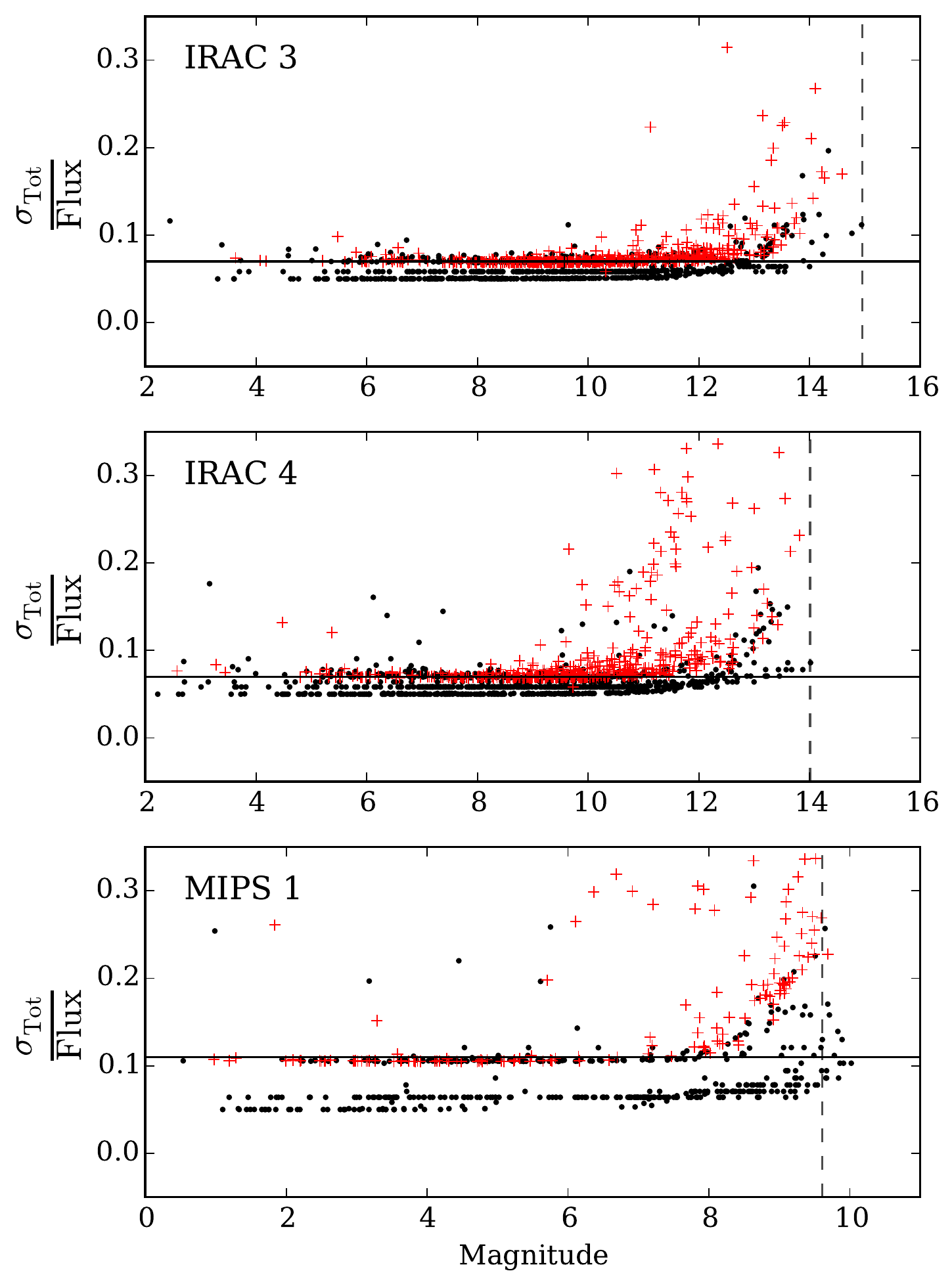} 
\end{figure}

We updated the study in R14 using new results from the literature as well as improved methods and calculations. In particular:

\begin{itemize}

\item the association $\epsilon$\,Cha was mistakenly named $\eta$\,Cha in R14;

\item  the distance to Serpens has been updated to 415$\pm$15 \,pc from Very Long Baseline Array observations of one star \citep{Bzid2010};

\item we have improved our estimate of the sensitivity completeness limits for each object by taking into account the extinction. The sensitivity limit should be compared against the reddened photospheric value. Reddened photospheres are flatter (once normalized) than derredened ones, and hence impose looser conditions for the detection limit. As a result, the total number of sources within the completeness limits increased significantly, allowing for better statistics and slightly varying the disk fractions in R14 (see Appendix);

\item the disk identification method in R14 is based on excess significances, which depend on the photometric errors (both observational and systematic). Hence, the disk fractions derived in our study depend strongly on the sensitivity of the {\it Spitzer} and WISE observations.  Smaller errors would result in a higher disk detection rate. The photometric data in R14 were compiled from different programs that used different techniques to estimate the photometric errors. Homogenizing the errors for the whole sample is therefore essential to guarantee a constant and coherent disk detection efficiency in our heterogeneous dataset. Figure \ref{fig:unc_dist} shows the photometric uncertainties as a function of apparent luminosity in IRAC3 and 4 and MIPS1 bands for the sample of 13 star-forming regions with {\it Spitzer} data. All regions but two (IC348 and Ophiuchus) have uncertainties at or below 7\% in all IRAC 3 and 4 bands, and at or below 12\% in MIPS1. We interpret the larger uncertainties for IC348 and Ophiuchus as the result of the conspicuous and highly variable mid-IR background observed in both regions at these wavelengths. To ensure a homogeneous disk detection rate across the entire sample, we rejected Ophiuchus and IC348 for the rest of our analysis, and set a conservative constant photometric error of 7\% for IRAC and WISE1, 2, 3 bands, and 12\% for MIPS1 and WISE4 in the $\chi_{\lambda}$ calculations. The final sample is comprised of 1\,809 sources (see Table~\ref{tab:diskfractions_previous}).

\end{itemize}

After these updates, we repeated the analysis in R14 and derived disk fractions in each association. We then fit these disk fractions as a function of time using an exponential law of the form:
\begin{equation}\nonumber
A e^{-t/\tau} + C
\end{equation}
where $t$ is the age of the region (in Myr), and $A$, $\tau$ and $C$ are left as free parameters. In this simple parametric model, $A$ represents the initial fraction of sources with IR excess, $\tau$ can be interpreted as the characteristic timescale of IR excess decay, and $C$ is a possible constant level.

The updated values are presented in the Appendix (Tables~\ref{tab:diskfractions_previous} and~\ref{tab:pp_debris_previous}, and Figs.~\ref{fig:diskfractions_previous} and~\ref{fig:pp_debris_previous}). These results are in agreement with previous studies \citep[e.g.][]{Lada2000, Hernandez2007, Hernandez2008, Murphy2013} and supersede our results in R14.

We note that we have combined data for regions at different distances, and therefore the sensitivity limit in spectral type is different for each of them. The overall completeness (all regions combined) for results with near-IR data (IRAC-WISE1,2,3) is around 0.1\,M$_\odot$. For mid-IR (MIPS1-WISE4), the sensitivity and the completeness limit increases to 0.4\,M$_\odot$.

\subsection{Impact of new membership estimates and region ages}\label{sec:members_and_ages}

New membership estimates are available in the literature for some of the regions and associations in this study \citep[e.g. $\epsilon$\,Cha, TW Hya][]{Murphy2013,Ducourant2014}, which could have an impact on our results. However, these works have modified the overall members of these associations by a small factor and moreover, our results will be still valid as long as the considered sample is representative of their stellar population. In addition, the original SACY sample has a homogeneous sensitivity for each region. By including other surveys with completely different selection criteria we might add systematics and biases difficult to detect and account for. Hence, for consistency with R14, we do not update membership or include new objects in this work.

Furthermore, the age estimates for the regions in our study are crucial for our analysis. Absolute ages of young associations are hard to determine, and some recent works have shown that the adopted values may be underestimated by a factor of two \citep{Bell2013}. However, the relative age sequence is easier to derive and is expected to be more precise \citep{Soderblom2013}. This is actually reinforced by the disk fractions derived for these regions: these measurements are independent of the age of the regions, yet they clearly decrease once sorted by age (see Appendix). We therefore assume the same age values as in R14, and we note that our estimates of disk lifetimes may be subject to modifications by a scaling factor, depending on the real absolute ages.

\section{Stellar masses and ages}\label{sec:masses_and_ages}

In this section, we study a possible dependence of disk evolution - as probed by disk fractions - as a function of stellar mass. To optimize the disk detection rate, we chose to limit the study to regions observed with the {\it Spitzer} Space Telescope. Its sensitivity and spatial resolution are significantly better than those of WISE, and allow a search for fainter excesses over the photospheric emission. A total of 1\,428 sources from 11 associations were observed with {\it Spitzer} (see Table~\ref{tab:regions_mass}). We note that this value is smaller than the initial sample, since we are not including associations studied with WISE data in R14 (i.e., the SACY sample). The better sensitivity and spatial resolution of the \textit{Spitzer} observations allow us to lower the threshold for excess detection. In the following we consider as excess any photometric measurements displaying $\chi \ge 3$ (instead of $\chi \ge 5$ used in Sect.~\ref{sec:previous_sample}).

\begin{table*}
  \caption{Young nearby star-forming regions included in the study of protoplanetary disk dependence on stellar mass. }
  \label{tab:regions_mass}
  \begin{center}
    \begin{tabular}{lccccc}
      \hline\hline Name & Age & Distance  & Membership and SpT & \textit{Spitzer} photometry & Number of sources\\
      & (Myr) & (pc)        &              &   &       \\
      \hline
      25~Orionis & 7-10 & 330 & (1) & (1) & 46\\
      Cha~I  & 2 & 160-165 & (2) (3) (4) & (3) (4) & 212\\
      Cha~II & 2$\pm$2& 178$\pm$18 & (5) & (31) & 47\\
      CrA     & 1-3 & 138$\pm$16 & (6) (7) (8) (9) & (32) & 35\\
      $\lambda$ Orionis  &  4 & 400$\pm$40 & (10) & (33) & 114\\
      Lupus  & 1-1.5 &140 - 200 & (11) (12) (13) (14) & (31) & 217\\
      NGC~1333  & 1 & 235$\pm$18 & (15) & (34) & 74\\
      $\sigma$ Orionis  & 2-3 & 440$\pm$30 & (16) (17) (18) (19) (20) (21) (22) & (35) & 104\\
      Serpens & 2 & 415$\pm$15 &  (15) (23) & (31) & 142\\
      Taurus & 1-2 & 140 &  (24) (25) (26) (27) & (26) (27) & 265\\
      Upper Sco & 11$\pm$2 & 140 & (28) (29) (30) & (29) & 405\\
      \hline
    \end{tabular}

    \tablebib{
      (1) \citet{Hernandez2007_25Ori}; 
      (2) \citet{Luhman2007}; 
      (3) \citet{Luhman2008a}; 
      (4) \citet{Luhman2008b}; 
      (5) \citet{Spezzi2008}; 
      (6) \citet{Neuhauser2000}; 
      (7) \citet{Nisini2005}; 
      (8) \citet{SiciliaAguilar2008}; 
      (9) \citet{SiciliaAguilar2011}; 
      (10) \citet{Bayo2011}; 
      (11) \citet{Krautter1997}; 
      (12) \citet{Allen2007}; 
      (13) \citet{Comeron2008}; 
      (14) \citet{Mortier2011}; 
      (15) \citet{Winston2009}; 
      (16) \citet{ZapateroOsorio2002}; 
      (17) \citet{Muzerolle2003}; 
      (18) \citet{Barrado2003}; 
      (19) \citet{Franciosini2006}; 
      (20) \citet{Caballero2007AA466}; 
      (21) \citet{Sacco2008}; 
      (22) \citet{Rigliaco2012}; 
      (23) \citet{Oliveira2009}; 
      (24) \citet{Luhman2004Taurus}; 
      (25) \citet{Monin2010}; 
      (26) \citet{Rebull2010}; 
      (27) \citet{Rebull2011}; 
      (28) \citet{Preibisch2002}; 
      (29) \citet{Carpenter2006}; 
      (30) \citet{Lodieu2011}; 
      (31) \citet{Evans2009};
      (32) \citet{Peterson2011}; 
      (33) \citet{Barrado2007}; 
      (34) \citet{Gutermuth2008}; 
      (35) \citet{Luhman2008SOri} 
    }

  \end{center}
\end{table*}

Stellar luminosities were obtained by scaling the best-fit photospheric model \citep[BT-Settl models,][]{Allard2012} to the dereddened photometric data (see R14) and integrating it over the entire stellar spectrum. In this way, the calculation does not depend on uncertain bolometric corrections. These luminosities were transformed into masses using the mass-luminosity relationship given by the pre-main sequence isochrones of the corresponding age in \citet{Bressan2012}. 

To study the influence of stellar mass in the evolution of protoplanetary disks, we divided the sample into two age bins and two mass bins:
\begin{itemize}

\item The age cut separating young and old objects was set to 3\,Myr. This is similar to the typical timescale of disk decay found in Sect.\ref{sec:previous_sample} (see Appendix) for protoplanetary disks.
A total of three regions (25 Ori, $\lambda$ Ori, and Upper Sco)  make up the ``old'' sample adding up to 489 sources, whereas the rest of the associations (939 sources) make up the ``young'' samplw.

\item The mass cut between low- and high-mass objects was set to 2\,M$_{\odot}$. Although somewhat arbitrary, this threshold is usually considered as the separation between T Tauri and Herbig stars. Given their different physical properties \citep[e.g. radiation fields, stellar winds or accretion rates, see ][]{Calvet2005, GarciaLopez2006,Hillenbrand2008}, it is likely that disks evolve differently around them. In addition, such a mass cut has the advantage of separating the sample so that both bins have enough sources for meaningful statistics.

\end{itemize}

Therefore, for the rest of the study we classified sources as:
\begin{itemize}
\item ``young'' if  1\,Myr $\leq$  age $\leq$ 3\,Myr, ``old'' otherwise
\item ``low-mass'' if M$_*$ $<$  2\,M$_{\odot}$, ``high-mass'' otherwise
\end{itemize}

\section{Disk fractions vs stellar mass and time}\label{sec:results}

\begin{figure*}
\caption{Evolution of protoplanetary disks (red) and processed (yellow) disks for each mass and age bin. The number of sources in each bin is shown within brackets. Errors are less than 6\,\% (see Table~\ref{tab:pie_charts_sum}).}\label{fig:piecharts_protoplanetary}
\centering
\includegraphics[width=0.85\textwidth]{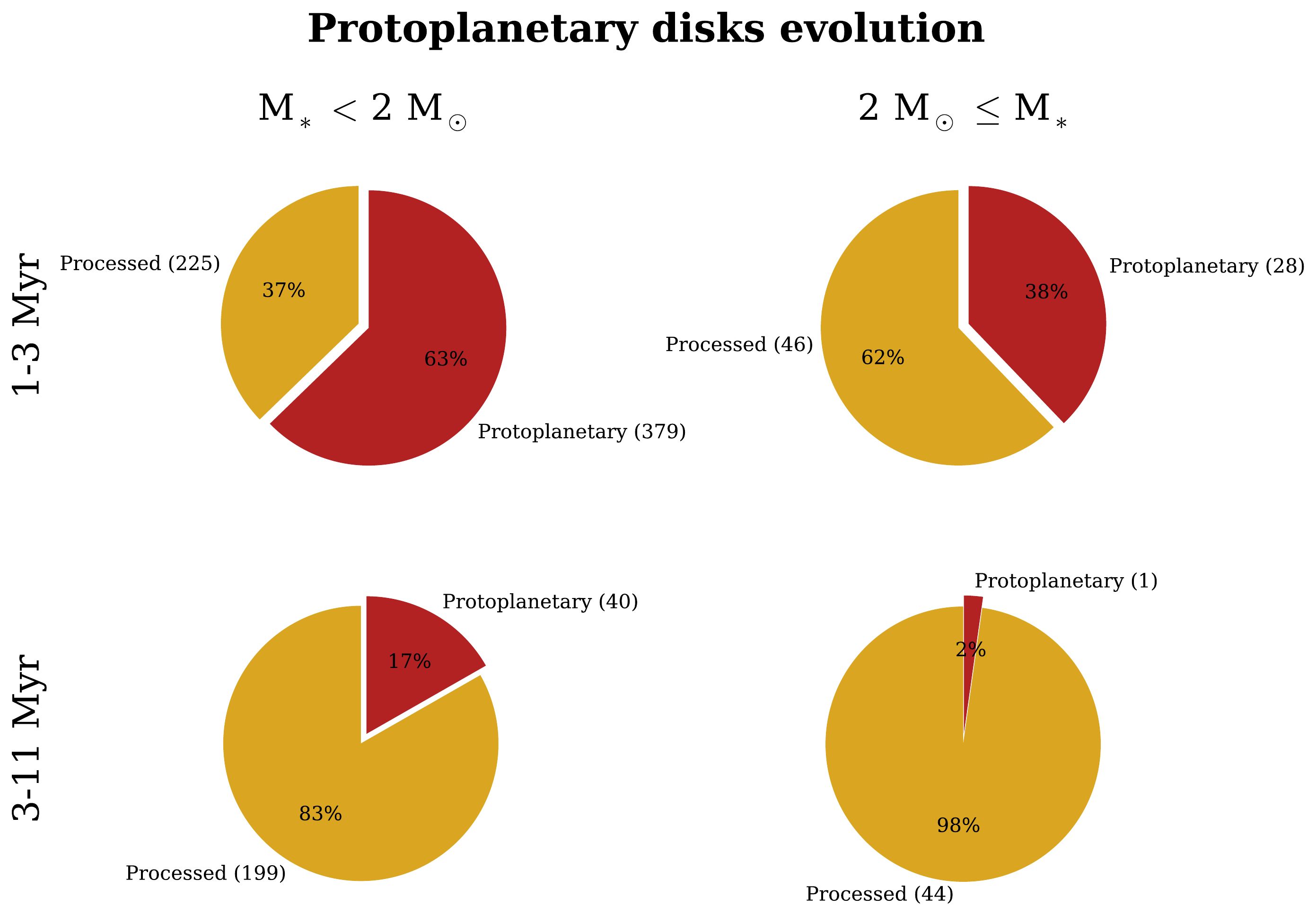}
\end{figure*}

The presence of dust around stars is usually detected as IR excess over the photospheric level: the dust is heated by the stellar radiation, and re-radiates with typical temperatures of tens or hundreds of Kelvins. However, the amount and shape of this excess depends strongly on the disk characteristics. Protoplanetary disks display excess at short-IR wavelengths \citep[$<$\,10\,$\mu$m, ][]{Williams2011}: these are disks that extend from the close regions ($<$ 1\,AU) of the stellar system to hundreds of AUs. However, the evolution of disks is thought to remove dust from the inner regions \citep[e.g.][R14]{Koepferl2012,Alexander2013}, leaving only mid-IR (24\,$\mu$m) or even far-IR ($>$\,50\,$\mu$m) excesses as in the case of transitional and debris disks \citep{Wyatt2008,Williams2011}.

\subsection{Protoplanetary disks}\label{sec:protoplanetary_disks}
In this section, we limit the study to short-IR excesses observed with IRAC, which offers the best sensitivity to the coolest objects of our sample, and therefore allows us to estimate disk fractions over a larger mass range. Unfortunately, IRAC photometry alone does not allow a fine classification of disks. We therefore define two simple groups: 
\begin{itemize}
\item protoplanetary disks are defined here as those sources having excesses at wavelengths shorter than 10\,$\mu$m (either 5.8 or 8\,$\mu$m, i.e., IRAC3 and/or IRAC4 bands). This definition encompasses the classic protoplanetary disks. 
\item processed disks are defined as sources without excess either at IRAC3 or IRAC4 bands. They could be transitional disks, debris disks, or diskless stars, but are indistinguishable based on IRAC photometry alone. In all cases their disks (if present) have reached a more advanced level of processing than the protoplanetary disks defined above.
\end{itemize}

We applied the sensitivity completeness correction as defined in Sect. 2 to ensure that the sample is not biased towards disk-bearing sources. A total of 1\,033 objects met this criteria for IRAC4 (the strictest sensitivity limit for IRAC bands). However, as pointed in Sect.\,\ref{sec:previous_sample}, the considered regions are at different distances and so the sensitivity limit translates into a different completeness limit in spectral types for each of them. We computed the limiting spectral type for each region, i.e. the latest spectral type detected with the corresponding sensitivity limit (see Table.\,\ref{tab:sensitivity}). Given its distance and age, 25 Orionis has a significantly earlier limiting spectral type than the rest of the regions, and we do not consider it in the analysis. Therefore, we cut the sample to the most restrictive limiting spectral type (M7, Cha II and $\lambda$ Orionis). After this procedure, the resulting IRAC sample is comprised of 962 sources.

We then divided this sample using the mass and age cuts described in Sect.~\ref{sec:masses_and_ages}. The result of this process is shown in the form of pie charts in Fig.~\ref{fig:piecharts_protoplanetary} and in Table~\ref{tab:pie_charts_sum}. Errors for the disk fractions were estimated using the bootstraping method further described in Sect.~\ref{sec:pdfs}, and are typically less than 6\,\%.

After inspection of Fig.~\ref{fig:piecharts_protoplanetary} some conclusions can be drawn:

\begin{itemize}

\item Protoplanetary disks are significantly more frequent around low-mass stars than around high-mass stars, independent of the age.

\item As expected, protoplanetary disk fractions decrease significantly at older ages for both low- and high-mass stars.

\end{itemize}

\subsection{Evolved disks}\label{sec:evolved}

As mentioned in Sect.~\ref{sec:protoplanetary_disks}, IRAC photometry alone does not allow a fine classification of processed disks. Photometric measurements at longer wavelengths are required to assess whether a source has a transitional or debris disk or if it is a diskless star. In an attempt to distinguish these categories, we repeated the analysis presented in Sect.~\ref{sec:protoplanetary_disks} including MIPS1 data. The 24~$\mu$m photometry allows he processed disks to be divided into two more precise categories:

\begin{itemize}

\item evolved disks, having no excess at IRAC3 or IRAC4, but with an excess at MIPS1. They include transitional and hot debris disks, as well as a small fraction of edge-on disks \citep[e.g][]{Huelamo2010} and circumbinary disks \citep[e.g][]{Ireland2008}.

\item diskless stars, displaying no excess at IRAC and MIPS1 bands. They include genuine diskless stars as well as colder (more evolved) debris disks that emit only at far-IR wavelengths.

\end{itemize}

The same special care was taken to avoid any sensitivity bias, and we kept only the sources with predicted photospheric fluxes above the sensitivity limit of the observations. MIPS1 observations are usually less sensitive than IRAC ones. We applied the same procedure as in the previous case and excluded the regions with significantly worse limiting spectral type ($\sigma$ Orionis). The final sample was limited to spectral types earlier than M4 (the corresponding limiting spectral type of CrA).  As a result, the total number of sources meeting this criteria dropped to 389. While smaller and encompassing a narrower mass range, this sample allows a more detailed look at the evolution of the inner disk via the study of transitional (evolved) disks.

\begin{figure*}
\caption{Frequencies for protoplanetary (red), evolved (blue) and diskless (gray) objects. The number of sources in each bin is shown in brackets. Errors are less than 8\,\% (see Table~\ref{tab:pie_charts_sum}).}\label{fig:piecharts_evolved}
\centering
\includegraphics[width=0.85\textwidth]{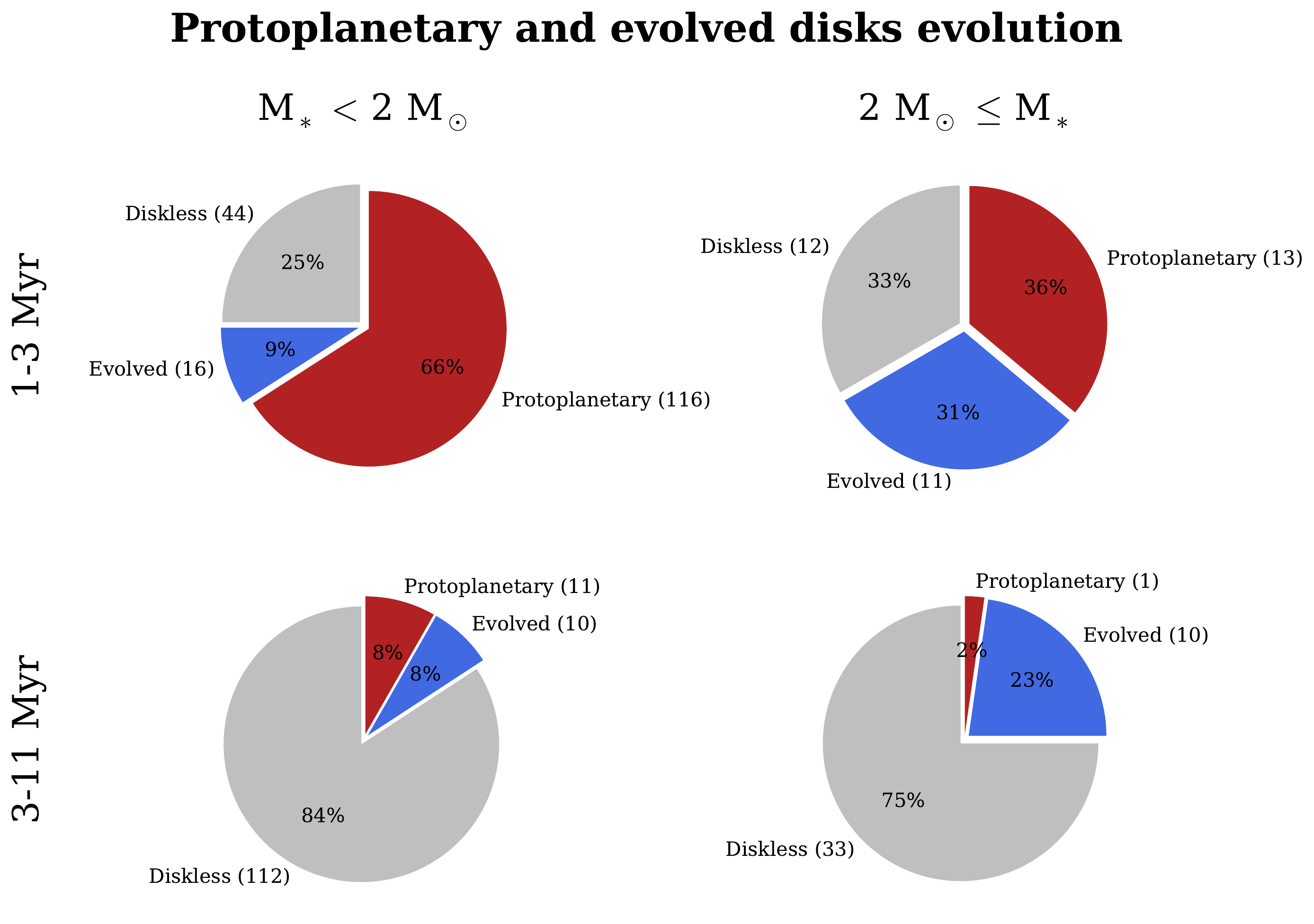}
\end{figure*}

The sample was then split into the mass and age bins defined previously. The results are shown in Fig.~\ref{fig:piecharts_evolved} and Table~\ref{tab:pie_charts_sum}. We note in particular that

\begin{itemize}

\item the results are consistent with those obtained in Section~\ref{sec:protoplanetary_disks} for protoplanetary disks within the uncertainties (see Sect.~\ref{sec:pdfs});

\item evolved disks are more frequent around high-mass stars than around low-mass stars, independently of the age;

\item a roughly constant fraction of evolved disks is found at young and old ages. A small difference seems to arise for high-mass stars (23\,\% to 31\,\%), but it is not statistically significant.

\end{itemize}

 \begin{table}
\small
  \caption{Sensitivity limits in spectral type for the IRAC and MIPS1 samples for each association considered in this study. Ellipses indicate that there are no objects in the sample with IRAC/MIPS1 above the corresponding sensitivity limit in that region. The adopted limiting spectral type for each case is marked with an asterisk.  Regions not included in the corresponding sample are specified.  }\label{tab:sensitivity}
  \begin{center}
    \begin{tabular}{lcc}
      \hline\hline
Region & IRAC sample & MIPS1 sample \\ 
\hline
25 Orionis & M5 (excluded) & \ldots \\
Cha I & M9 & M6 \\
ChaII & M7* & M5 \\
CrA & M8 & M4* \\
$\lambda$\,Orionis & M7* & \ldots \\
Lupus & M9 & M9 \\
NGC1333 & M8 & M5 \\
$\sigma$\,Orionis & M8 & M0 (excluded) \\
Serpens & M9 & M8 \\
Taurus & M9 & M6 \\
Upper Sco & M9 & M5 \\
\hline
\end{tabular}
    \end{center}
\end{table}

\subsection{Significance of the results}\label{sec:pdfs}

The large number of sources included in the various samples in Sects.~\ref{sec:protoplanetary_disks} and \ref{sec:evolved} does not guarantee the significance of the differences observed in Figs.~\ref{fig:piecharts_protoplanetary} and~\ref{fig:piecharts_evolved} and Table~\ref{tab:pie_charts_sum}. Calculating uncertainties using standard error propagation or Poisson statistics is neither practical nor trivial in the case of our analysis. The computed fractions depend on several different parameters (e.g., photometric errors, $\ T_{eff}$ value and $A_V$ fit, photospheric models and adopted isochrones,...) and the age and mass thresholds selected to split the sample could have an impact on the results and their interpretation. 

For these reasons, we tentatively estimate uncertainties on the disk fractions by performing bootstrapping (1000 iterations) randomly varying the $T_{\rm eff}$, photometric fluxes, the young/old boundary, the high mass/low mass boundary, and the object distances within normal distributions with $\sigma$ of 50\,K, $\sigma_{flux}$, 0.5\,Myr, 0.25\,M$_{\odot}$, and error in the region's distance (or 20\,pc when no error was available). Uncertainties on the age of the sources were not included in the bootstrapping; although absolute ages are hard to determine, relative ages are expected to be more accurate \citep{Soderblom2013}. Additionally, our analysis considers only two clearly separated age bins.

The bootstrapping yielded probability density functions (PDFs) of disk fraction for each sample, and should account for most uncertainties in the calculations. The standard deviation of the PDFs (which are close to normal) provides a reasonable estimate of the final uncertainties on the disk fractions and are reported in Table~\ref{tab:pie_charts_sum}.

Overall, the estimated uncertainties suggest that the differences seen in the disk fractions are statistically significant, and that the distributions for high-mass and low-mass stars are different.

We also checked our results to vary little (a few percentage points) when modifying the mass cut between 1.5 and 2.5\,M$_\odot$, and hence our conclusions are robust against the selected mass threshold. Given the lack of regions in the 4-8\,Myr regime, it is not possible to test the impact of a different age cut in detail. Our aim is to separate sources which are clearly younger or older, and this value is likely not to have a very strong impact on our study. For this reason, our results are weakly dependent on age uncertainties, as long as relative ages are properly estimated (see discussion in Sect.~\ref{sec:members_and_ages}).

\begin{table*}
\small
  \caption{Results from the analysis of disk frequencies as a function of age and mass. See also Figs.~\ref{fig:piecharts_protoplanetary} and~\ref{fig:piecharts_evolved}. Errors are one standard deviation from the computed PDFs.}\label{tab:pie_charts_sum}
  \begin{center}
    \begin{tabular}{lcccc}
      \hline\hline
Disk type & Young+Low-mass [\%] & Young+High-mass [\%] & Old+Low-mass [\%] & Old+High-mass [\%] \\ 
\hline
      \multicolumn{5}{c}{Protoplanetary disks (Section~\ref{sec:protoplanetary_disks})}\\ 
      \hline
Protoplanetary (IRAC4 detection limit) & 63$\pm$2 & 38$\pm$6 & 17$\pm$2 & $2^{+4}_{-2}$\\
Processed (IRAC4 detection limit) & 37$\pm$2 & 62$\pm$6 & 83$\pm$2 & $98^{+2}_{-4}$\\
\hline
      \multicolumn{5}{c}{Protoplanetary and evolved disks (Section~\ref{sec:evolved})}\\ 
      \hline
Protoplanetary (MIPS1 detection limit) & 66$\pm$4 &  36$\pm$8 & 8$\pm$2 & $2^{+4}_{-2}$ \\ 
Evolved (MIPS1 detection limit) & 9$\pm$2 &  31$\pm$7 & 8$\pm$3 & 23$\pm$6 \\
Diskless (MIPS1 detection limit) & 25$\pm$3 & 33$\pm$8 & 84$\pm$3 & 75$\pm$7 \\
\hline
\end{tabular}
    \end{center}
\end{table*}

\subsection{Disk fractions and stellar temperature (mass)}

Our definition of protoplanetary, evolved and diskless sources depends on the wavelength at which excess over the photosphere is found. Because the temperature distribution in the circumstellar disk depends directly on the stellar effective temperature and luminosity, the classification probably suffers from a dependence on stellar temperature. 
Infrared emission at a certain wavelength likely originates from different regions of the disk for hot (massive) and cool (low-mass) stars, and may even arise from different disk types. 

This effect is probably not critical for our analysis of protoplanetary sources (Sect.~\ref{sec:protoplanetary_disks}). These dense and massive disks have complex physical structures, and the IR excess at a certain wavelength is emitted from relatively large areas of the disk. Moreover, the protoplanetary definition is based on the presence of excess over a relatively wide wavelength range, ensuring that the corresponding emission comes from broad areas of the disk regardless of the stellar mass. 

On the other hand, our definition of evolved disks relies primarily on IR excess at 24\,$\mu$m (MIPS1). The area of the disk probed at 24~$\mu$m is expected to vary significantly between massive and low-mass stars, and might make the comparison of the excess fractions and their interpretation more difficult. 

To test how much this dependence might affect our results, we computed a grid of radiative transfer models of circumstellar disks using the code MCFOST \citep{MCFOST}. We found that synthetic SEDs classified as evolved according to our criterion always corresponded to higher levels of disk processing regardless of the star's temperature. These simple simulations suggest that our classification scheme is efficient and robust enough to allow a comparison of the results obtained for high and low-mass stars.

\begin{figure}
\caption{Comparison of the obtained PDFs for evolved sources as a function of age and for the two stellar mass bins.}\label{fig:evolved_mass_comp}
\centering
\includegraphics[width=\hsize]{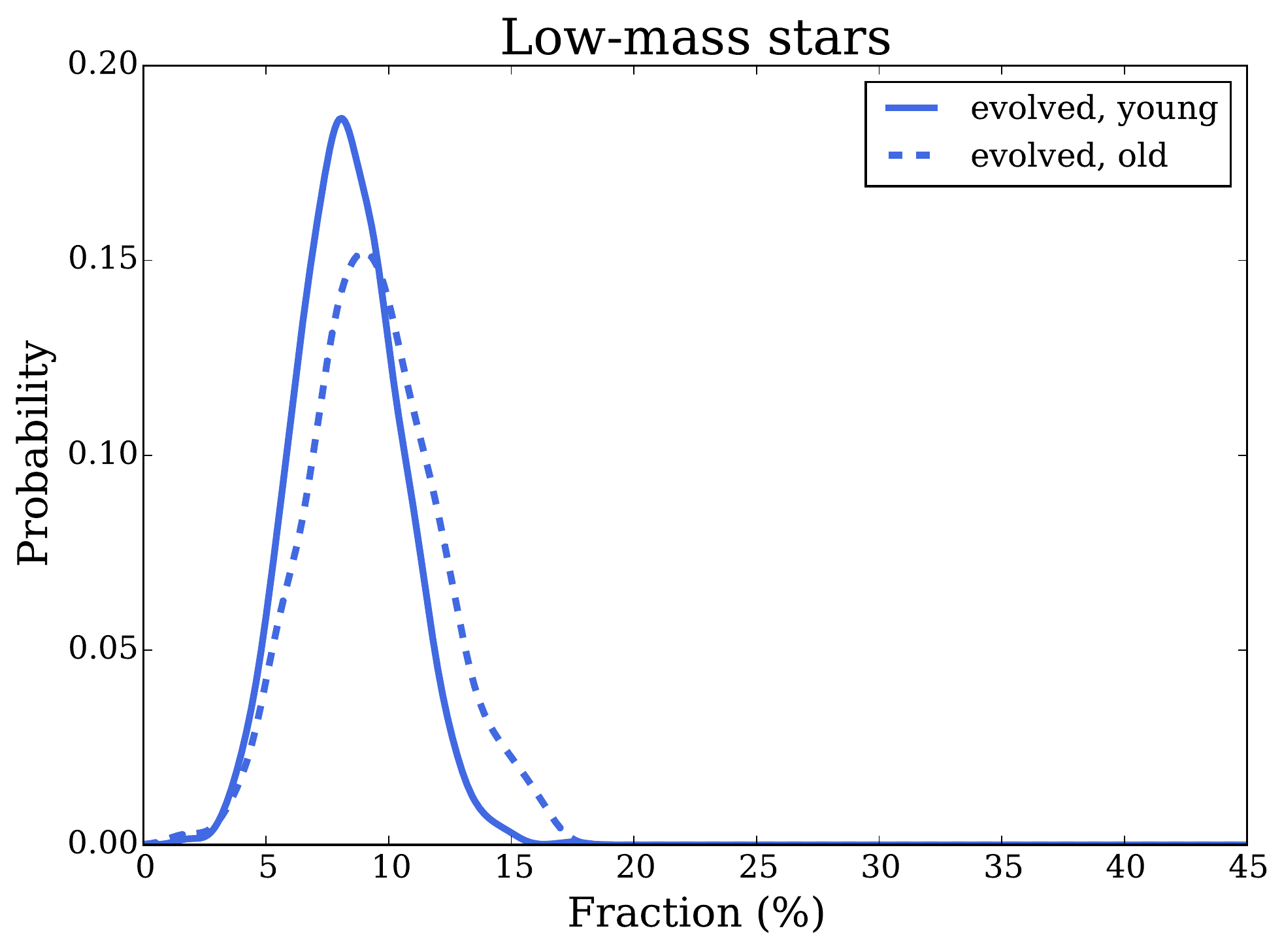}
\includegraphics[width=\hsize]{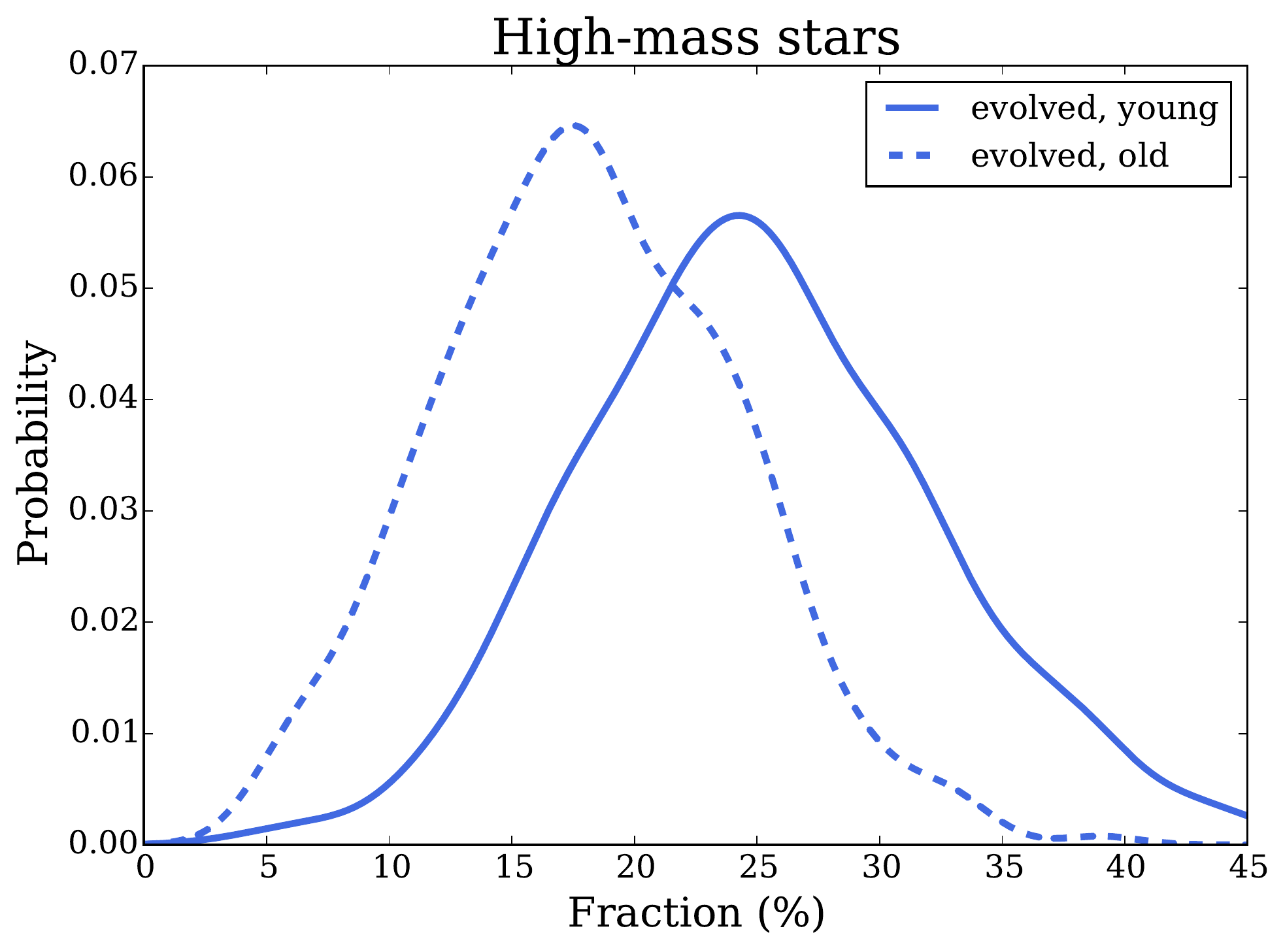}
\end{figure}

Finally, a comparison of the evolution of evolved disks can be performed within the same mass bin, removing most of the previous dependency on stellar temperature. Figure~\ref{fig:evolved_mass_comp} shows the PDFs of the "evolved" disk fractions obtained in Sect.~\ref{sec:pdfs} for high-mass stars on the one hand and low-mass stars on the other hand. The PDFs are completely compatible both for low-mass and high-mass stars, pointing to a roughly constant level of evolved disks within the uncertainties at any age regardless of stellar mass.

\section{Impact of stellar mass on disk evolution}\label{sec:discussion}

\subsection{Dependence of protoplanetary disk lifetimes on stellar mass.}

\begin{figure}
\caption{Evolution of protoplanetary disks as a function of mass, as derived in this study.}\label{fig:scheme}
\centering
\includegraphics[width=\hsize]{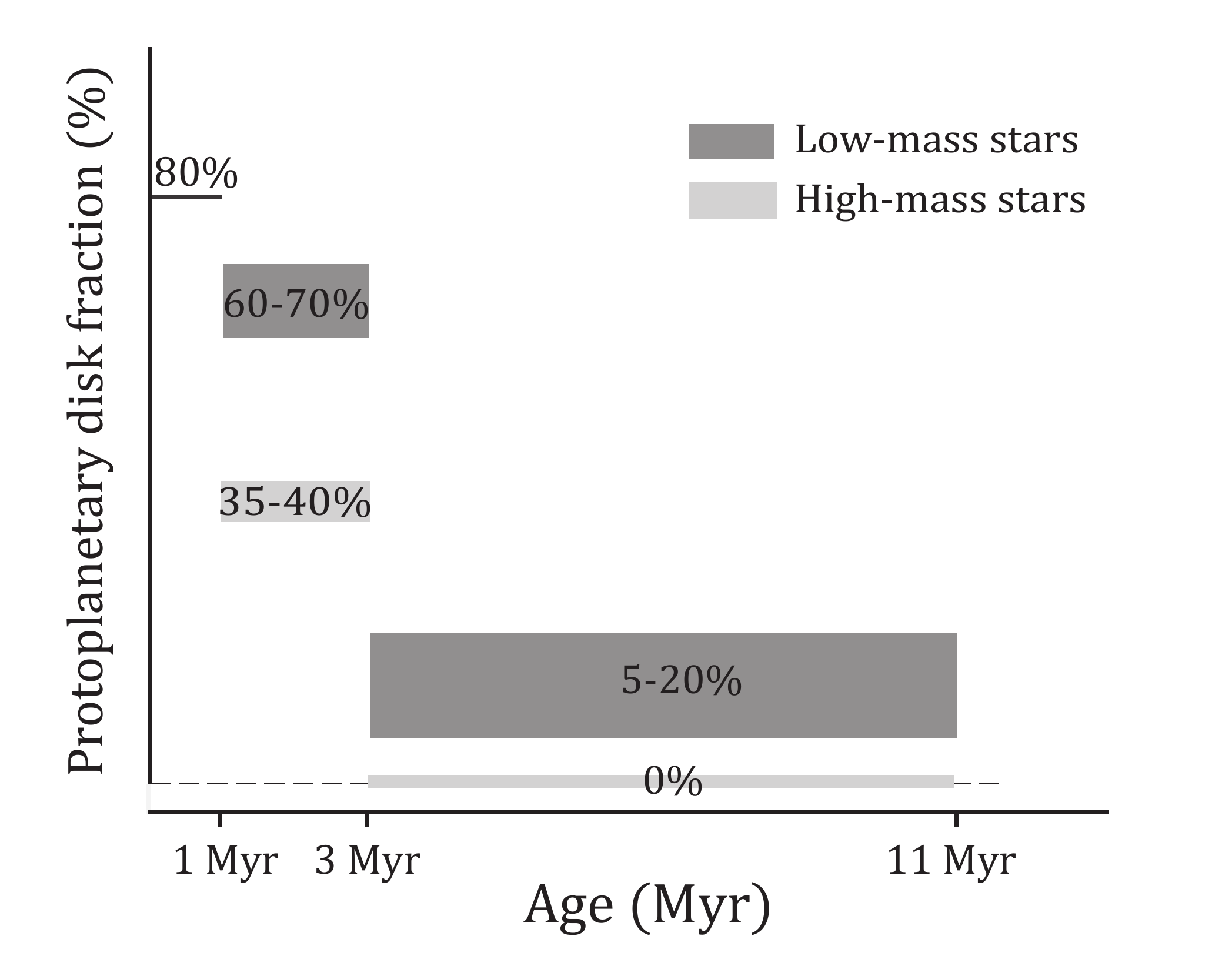}
\end{figure}

Previous observational studies had already found lower protoplanetary disk fractions around high-mass stars and suggested that their protoplanetary disks disperse earlier and/or evolve faster \citep[see e.g][for a review ]{Williams2011}. A number of these studies focused on individual regions, making it hard to confirm a global dependence of protoplanetary disk evolution on stellar mass independently of the environment and method used to identify the disks \citep[e.g][ in IC348, NGC~2362 and Upper Sco]{Lada2006,Dahm2007,Currie2009_NGC2362, Carpenter2006}.  \citet{Kennedy2009} performed an original and coherent study of disk frequencies in nine young associations by combining optical spectra and IR photometry. In contrast to our study, they found only a weak dependence of disk dispersal on stellar mass. The ages used in their study have been improved and sometimes changed significantly. For example, the age of Upper Sco was recently re-estimated to 11\,Myr \citet{Pecaut2012} from an original 5~Myr \citep{Preibisch1999}, resulting in a noisier distribution of disk fraction as a function of age in their study.

Our results demonstrate quantitatively that protoplanetary disks dissipate significantly faster and/or earlier around high-mass stars, as we find between 10 and 30\,\% more protoplanetary disks around low-mass stars regardless of the age. 

Our data additionally show that the disk fraction for high-mass stars falls to $\approx$0\% anytime between 3 and 11~Myr, and suggest that protoplanetary disks disperse up to two times faster around high-mass stars than around low-mass ones. More data for regions with intermediate ages are needed to better constrain this ratio.

Interestingly, other studies found a higher (80-85\,\%) protoplanetary disk fraction and no dependence with stellar mass in the younger Trapezium cluster \citep[$<$\,1\,Myr,][]{Lada2000} and NGC~2024 cluster  \citep[$<$\,0.4-1.4x10$^5$\,yr,][]{Haisch2000}. Combined with the present study, these results suggest that
\begin{itemize}
\item the disk fractions before 1~Myr are in the continuity of the results presented here;
\item the dependence of disk fractions on stellar mass appears after the first Myr of the star's lifetime, and persists until no protoplanetary disks are found around $\sim$\,10\,Myr
\end{itemize}
Based on these two suggestions, we propose a scheme of disk evolution as a function of stellar mass (see Fig.~\ref{fig:scheme}). The timescale of protoplanetary disk formation and evolution is mostly independent of stellar mass until the first Myr. At 3\,Myr, significant differences are already observable between the evolution of protoplanetary disks around low- and high-mass stars. By $\sim$\,10\,Myr, no protoplanetary disks are found around high-mass stars, but $\approx$10-15\,\% of low-mass stars still harbor a disk. Stronger radiation fields and higher accretion rates of high-mass stars \citep{Calvet2005, GarciaLopez2006,Hillenbrand2008} are likely to affect the evolution of protoplanetary disks even during the first Myrs, fastening the processes responsible of disk dispersal \citep{Alexander2013}.

\subsection{A constant level of evolved disks}\label{sec:evolved_disks}

We find the fraction of evolved sources to remain constant with age both for low- and high-mass stars: 5-15\,\% of low-mass stars display this kind of excess, and the fraction increases to 20-30\,\% for high-mass objects (see Fig.~\ref{fig:evolved_mass_comp}). Again, the dependence with stellar mass of the disk radii probed at a certain wavelength should be consider and it could, in fact, account for the higher fraction of evolved disks around high-mass stars. Nevertheless, the obtained values at these ages can be established to be within 5-30\,\%.

The evolved disk definition proposed here includes transitional disks, hot debris disks, edge-on disks and circumbinary disks. Unfortunately, the classification does not allow us to discriminate between these four types of disks. We nevertheless note that simple geometric considerations suggest that edge-on disks must be relatively rare contribute only moderately to the population of evolved disks presented here. Hot debris disks are expected to appear mostly at later stages of disk evolution and must populate mostly the older bin of our analysis. Circumbinary disks are expected to be common given the multiplicity frequency among stars \citep[e.g][]{Bouy2011}, but only the closest systems would mimic the SED of evolved disks defined here \citep[e.g as in ][]{Ireland2008,Nguyen2012}. The evolved disks in the young ($<$\,3\,Myr) bin are therefore most likely primarily made of transitional and circumbinary disks, and our analysis suggests that these two types of disks can be found even during the very first Myrs of stellar life.

Finally, we also note that the higher fraction of evolved disks -- which include circumbinary disks -- around high-mass stars is also consistent with their higher multiplicity frequency \citep{Duchene2013}.

\subsection{Transitional disks} 

We stress that the current data in our sample cannot be used to isolate transitional disks among evolved disks. With this limitation in mind, it is nevertheless interesting to compare the properties of evolved disks of the present study to the properties of transitional disks as reported in the literature.

Transitional disks are a particularly interesting kind of circumstellar disks that harbor a gap and/or cavity \citep{Andrews2011}. Various mechanisms have been proposed to explain the existence of these gaps and cavities. They can be produced by the formation of planets \citep{Espaillat2014}, grain growth, photoevaporation or the formation of low-mass companions \citep{Birnstiel2012,Alexander2013,Papaloizou2007}. The connection between transitional disks and planets -- as demonstrated by several resolved images of planetary mass companions within the gap or cavity \citep[e.g][]{Huelamo2011,Kraus2012,Quanz2013,Close2014} -- makes them a cornerstone for theories of planet formation and evolution. 

First, we note that the fraction of evolved disks derived by our analysis (between 5 and 30\%) is consistent with transitional disk fractions reported in the literature \citep[e.g.][]{Skrutskie1990, Kenyon1995, Andrews2005, Cieza2007, Merin2010}. 

Second, previous studies have established the duration of the transition phase to be $\approx$\,1\,Myr \citep{Alexander2013}, which is in agreement with the combined 5-30\,\% of evolved sources and the $\approx$10\,Myr disk lifetime, if one assumes that all disks go through a transitional phase. Within the planet formation scenario, only Jovian planets are likely to produce a gap in their protoplanetary disk \citep{Lin1986,Marsh1992,Nelson2000,Calvet2002,Rice2003}, and so it is also plausible that not all planetary systems go through a transitional phase during their formation. If that is the case, then the estimated $\approx$\,1\,Myr duration of this phase should be regarded as a lower limit.

\subsection{Implications for planet formation and giant planet migration}

To date, there are two main theories explaining the formation of planets: the core accretion model (CA) and the disk instability model (DI) \citep[see][for recent reviews on this topic]{Helled2013,Raymond2013}:
\begin{itemize}

\item In the CA scenario, a solid planetary embryo is formed by continuum accretion of planetesimals onto this embryo. Terrestrial planets are likely formed in this way.  If the solid core becomes massive enough, it will also start accreting gas at an increasing rate until no more gas can be obtained from the disk (either because the disk has dissipated or because the planet has opened a gap in it). The result of this process is a gas giant planet. 

\item In the DI model, a very massive disk is destabilized by its own gravity, leading to the collapse of a region of the disk and potential formation of a giant planet or brown dwarf. 

\end{itemize}

So far, the CA model seems to better reproduce several observed properties of the known exoplanets. For example, correlations are known to exist, both between the stellar metallicity and giant planet occurrence, and between stellar and planetary metallicities \citep[e.g.][]{Johnson2010}. These correlations can be explained by the CA model, in which higher metallicities result in more efficient planet formation. However, some known exoplanetary systems cannot be explained within the CA frame \citep[e.g. the system around HR 8799,][]{Marois2008}, suggesting that DI may be responsible for the formation of some gas giants at very large radii from their host stars and/or around low-metallicity stars. It could also explain the existence of some brown dwarfs orbiting other stars. The global planet formation process may be a combination of these two scenarios.

If CA is the dominant mechanism, then planets are likely to form below the ice-line \citep{Ida2004a}: water can freeze into ice around the grains of dust, and increase the efficiency of their aggregation to form the required planetary embryo. On the other hand, formation at very large radii from the host star is also unlikely, since the solid accretion rate of planetary embryos decreases with the distance to the star. As an example, the optimal zone to form planets for a solar-type star is estimated to be between 5-10\,AU \citep{Helled2013}. Surprisingly, a significant number of Jovian-like planets are found very close to their stars, and are referred to as hot Jupiters. This particular subgroup of exoplanets suggests that a fraction of these gas giants migrate inwards after their formation: once a gap is opened by the planet, it gets locked to the disk and exchanges angular momentum with it, causing the planet to move closer to its host star \citep[Type-II migration,][]{Papaloizou2007}. Other migration processes such as planet-planet scattering, Kozai resonances, or similar dynamic mechanisms \citep[e.g.][]{Kozai1962,Wu2003,Naoz2011} are likely responsible for the fraction of hot Jupiters whose spin is found to be misaligned with their planet-star orbits \citep{Triaud2010,Narita2010,Morton2011}.

Within this theoretical framework, the obtained dependence of stellar mass and disk lifetime would have two main implications:

\begin{enumerate}

\item Observational results tentatively suggest that massive stars are more likely to host gas giants than their low-mass counterparts \citep{Johnson2010}. This trend would be the result of two competing factors: high-mass stars have more massive disks \citep[e.g.][]{Andrews2013}, and so are more likely to form gas giants. On the other hand, the present study shows that massive stars disperse their disks earlier. Under some circumstances, the formation of giant planets could even last for some Myr \citep{Helled2013}. Therefore, a fast dispersing disk may not live long enough to form gas giants. If the observed trend (massive stars being more likely to host giant planets) is real, then either a strong correlation between the stellar mass and the ability to form giant planets exists to compensate for the shorter disk lifetime, or giant planet formation is faster than the disk dispersal time regardless of the stellar mass, or most of the planet formation process happens in the first million years.

\item Type-II migration is active while the gaseous disk is present and the planet can exchange angular momentum with it. The distribution of orbital semi-major axes produced by these migration mechanisms is controlled by the ratio of the disk depletion timescale to the viscous diffusion timescale at the distance where Jovian planets form \citep{Armitage2002,Trilling2002,Ida2004b}. A faster dispersal of protoplanetary disks around high-mass stars would result in a shorter disk depletion timescale, and therefore a shorter migration period \citep{Burkert2007,Currie2009_hotjupiters}. Our results provide a robust confirmation of a faster disk dispersal around high-mass stars: the disk lifetime could be more than two times shorter for massive stars, reducing significantly the migration period available for Jovian planets and, as a result, a smaller number of hot Jupiters would be present around these stars.

\end{enumerate}

\begin{figure}
\caption{Orbital semi-major axes as a function of the mass of the host star for the known exoplanet population. Red circles show hot Jupiters (M $>$ 1\,M$_{J}$, period $<$ 10\,days), black circles represent the rest of the sample. The size of the symbols scales with the planet mass.  We also show the estimated stellar mass radii of main-sequence stars \citep{Siess2000} and the stellar radii of the host stars. A clear gap of hot Jupiters is seen for M$_{*} >$ 1.5\,M$_{\odot}$. }\label{fig:exop_distribution}
\centering
\includegraphics[width=\hsize]{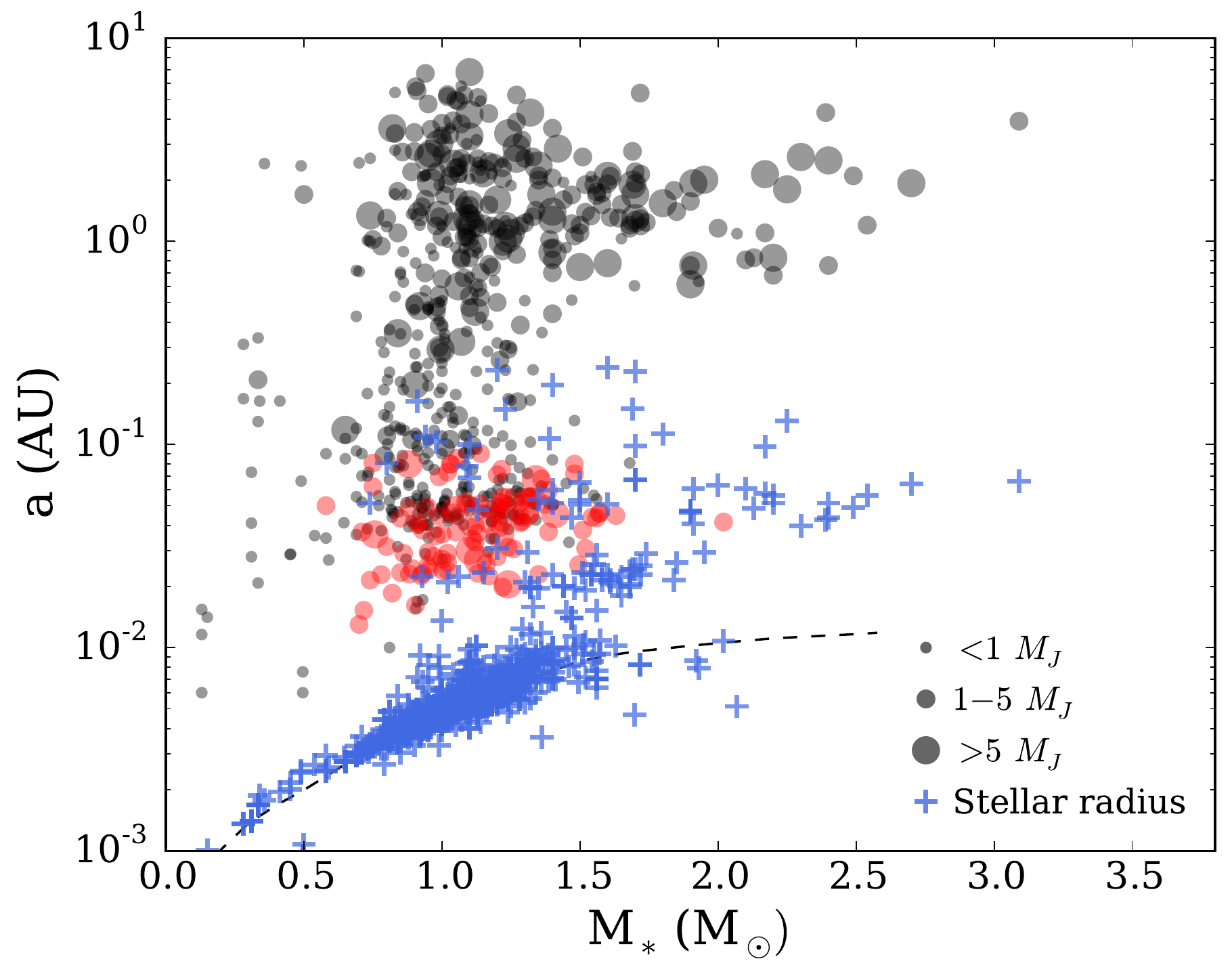}
\end{figure}

Following \citet{Kennedy2009}, point 2 can be linked with the known population of exoplanets. Figure~\ref{fig:exop_distribution} shows the orbital semi-major axis distribution as a function of the mass of the host star for the known exoplanets\footnote{as in the \url{www.exoplanets.eu} database, August 6$^{\rm th}$ 2014} (planets around pulsars are not plotted since they represent a completely different type of object). Considering as  hot Jupiters the planets with masses above 1\,M$_{J}$ and periods $<$\,10 days, a clear lack of these objects around massive stars ($>$\,1.5 M$_{\odot}$) is easily identified. This paucity of close Jovian planets around massive stars is still a matter of intense debate. Tidal interaction between massive planets and their host stars could lead to planet engulfment \citep{Villaver2009,Villaver2014}. Some authors report a real dependence on stellar mass \citep[e.g.][]{Johnson2011}, and have even shown hints of it not being the result of planet engulfing caused by the post-main-sequence evolution of the star \citep{Johnson2007}. On the other hand, other authors suggested that such claims were made on incorrectly assigned stellar masses \citep{Lloyd2011}, and that tidal destruction is responsible for the observed trend \citep{Schlaufman2013}. By also plotting the radii of the planet host stars in Fig.~\ref{fig:exop_distribution}, it becomes clear that most low-mass stars ($<\,1.5-2\,M_{\odot}$) have radii matching those predicted from main-sequence models \citep{Siess2000}. For more massive stars, the stellar radii increase owing to the post-main-sequence phase, overlapping with the loci of the hot Jupiters semi-major axes. It is not possible to tell if the massive stars in Fig.~\ref{fig:exop_distribution} have engulfed any close gas giants, but it becomes clear that the post-main-sequence phase of these stars could account for a fraction of the observed paucity of hot Jupiters around them. Assessing the origin of this lack of close gas giants is beyond the scope of this study, but the faster disk dissipation around high-mass stars could contribute to it as well.

This discussion may also apply to some brown dwarfs. Although it is thought that most of these objects form via different mechanisms \citep{Chabrier2014}, the existence of binary systems including a brown dwarf companion suggests that some of them may form in a circumstellar disk around more massive stars. If this is the case, they are very likely formed in similar ways to giant planets, and the same conclusions could be applied to this particular subgroup of brown dwarfs. 

Finally, the implications of the faster disk dispersal around high-mass stars for terrestrial planets are harder to determine. As previously mentioned, their formation follows the initial stage of the CA model, accreting planetesimals to form a planetary embryo and eventually a rocky planet, which will not become a gas giant if the planet cannot accrete enough gas for any reason. However, this process is faster than gas giant formation: earth-mass planets can be formed in $~$1\,Myr \citep{Raymond2013}, and so the disk dissipation dependence with stellar mass may not be as relevant as for giant planet formation. Moreover, terrestrial planets do not undergo Type-II migration, and therefore their orbits are less dependent on the disk properties. However, their orbits can be strongly affected by the migration of giant planets in the same system. If the migration history of gas giants has a dependence with stellar mass, it could indirectly produce different orbits for terrestrial planets too. The identification of any possible trend of this effect requires further modeling and study of the known planetary population.

\section{Conclusion}\label{conclusions}

We have studied the disk lifetime dependence on stellar mass using the sample of YSOs compiled by \citet{Ribas2014}. After updating their results, we have divided the sample into low- or high-mass stars (boundary set to 2\,M$_{\odot}$) and young or old regions (boundary set to 3\,Myr). We then study the fraction of protoplanetary and evolved disks as a function of age and mass. The large number of sources in the sample allows us to confirm the existence of a dependence of disk lifetime on the stellar mass: high-mass stars disperse their disks up to two times faster and/or earlier than low-mass ones. We also find that the fraction of evolved disks (including transitional, circumbinary, and hot debris disks) remains roughly constant (5-30\,\%) during the first $~$\,10\,Myr of the stellar life. The faster dispersal of protoplanetary disks around high-mass stars could have important implications for giant planet formation and migration, and may contribute to the apparent lack of hot Jupiters around these stars. 

\begin{acknowledgements}

This work has been possible thanks to the ESAC Science Operations Division research funds with code SC 1300016149, support from the ESAC Space Science Faculty and of the Herschel Science Centre. We thank the referee for valuable comments which have helped to improve the paper. We also acknowledge Gaspard Duch\^ene, Johannes Sahlmann, Jorge Lillo-Box,  Jos\'e Antonio Caballero, and Justin Crepp for helpful discussions on disks, Jovian planet populations, migration and biases in the sample of known exoplanets.
H. Bouy is funded by the Spanish Ram\'on y Cajal fellowship program number RYC-2009-04497. This work has made an extensive use of Topcat \citep[TOPCAT \url{http://www.star.bristol.ac.uk/~mbt/topcat/} and STILTS,][]{Topcat,Stilts}. This work is based in part on data obtained as part of the UKIRT Infrared Deep Sky Survey. This research made use of the SDSS-III catalog. Funding for SDSS-III has been provided by the Alfred P. Sloan Foundation, the Participating Institutions, the National Science Foundation, and the U.S. Department of Energy Office of Science. The SDSS-III web site is http://www.sdss3.org/. SDSS-III is managed by the Astrophysical Research Consortium for the Participating Institutions of the SDSS-III Collaboration including the University of Arizona, the Brazilian Participation Group, Brookhaven National Laboratory, University of Cambridge, University of Florida, the French Participation Group, the German Participation Group, the Instituto de Astrofisica de Canarias, the Michigan State/Notre Dame/JINA Participation Group, Johns Hopkins University, Lawrence Berkeley National Laboratory, Max Planck Institute for Astrophysics, New Mexico State University, New York University, Ohio State University, Pennsylvania State University,  University of Portsmouth, Princeton University, the Spanish Participation Group, University of Tokyo, University of Utah, Vanderbilt University, University of Virginia, University of Washington, and Yale University. 
This work makes use of data from the DENIS Survey. DENIS is the result of a joint effort involving human and financial contributions of several Institutes mostly located in Europe. It has been supported financially mainly by the French Institut National des Sciences de l'Univers, CNRS, and French Education Ministry, the European Southern Observatory, the State of Baden-Wuerttemberg, and the European Commission under networks of the SCIENCE and Human Capital and Mobility programs, the Landessternwarte, Heidelberg and Institut d'Astrophysique de Paris. 
This publication makes use of data products from the Wide-field Infrared Survey Explorer, which is a joint project of the University of California, Los Angeles, and the Jet Propulsion Laboratory/California Institute of Technology, funded by the National Aeronautics and Space Administration.

This research used the facilities of the Canadian Astronomy Data Centre operated by the National Research Council of Canada with the support of the Canadian Space Agency.   

This publication makes use of data products from the Two Micron All Sky Survey, which is a joint project of the University of Massachusetts and the Infrared Processing and Analysis Center/California Institute of Technology, funded by the National Aeronautics and Space Administration and the National Science Foundation.

This work is based in part on observations made with the Spitzer Space Telescope, which is operated by the Jet Propulsion Laboratory, California Institute of Technology under a contract with NASA.

This research has made use of the Exoplanet Orbit Database and the Exoplanet Data Explorer at exoplanets.org.
\end{acknowledgements}

\bibliographystyle{aa}
\bibliography{biblio}

\begin{appendix}
\section{Updated figures and tables from R14}
\begin{table*}
\caption{Fraction of sources with excess for each region in the three wavelength regimes (update of Table~4 in R14)}
\label{tab:diskfractions_previous}
\begin{center}
\begin{tabular}{lcccc}
\hline\hline Name & Fraction $_{\rm IRAC}$(\%) & Fraction $_{\rm short}$(\%) & Fraction $_{\rm intermediate}$(\%) & Fraction $_{\rm long}$(\%) \\
\hline
25 Orionis & 12 $\pm$ 7 [26] & 8 $\pm$ 5 [26] & 12 $\pm$ 7 [25] & \ldots [0]\\
Cha I & 51 $\pm$ 4 [117] & 36 $\pm$ 5 [109] & 52 $\pm$ 5 [106] & 71 $\pm$ 6 [52]\\
Cha II & 75 $\pm$ 8 [28] & 50 $\pm$ 9 [28] & 85 $\pm$ 9 [20] & 82 $\pm$ 7 [28]\\
CrA & 40 $\pm$ 10 [19] & 11 $\pm$ 8 [19] & 50 $\pm$ 10 [16] & \ldots [6]\\
IC 348 & 38 $\pm$ 3 [253] & 17 $\pm$ 2 [262] & 41 $\pm$ 3 [223] & 81 $\pm$ 8 [27]\\
$\lambda$-Orionis & 23 $\pm$ 7 [43] & 10 $\pm$ 4 [51] & 23 $\pm$ 6 [43] & \ldots [0]\\
Lupus & 53 $\pm$ 5 [85] & 27 $\pm$ 5 [86] & 53 $\pm$ 5 [85] & 57 $\pm$ 6 [58]\\
NGC 1333 & 64 $\pm$ 5 [73] & 38 $\pm$ 6 [73] & 67 $\pm$ 6 [70] & 88 $\pm$ 8 [17]\\
Ophiuchus & 23 $\pm$ 3 [248] & 13 $\pm$ 2 [253] & 23 $\pm$ 3 [238] & 42 $\pm$ 5 [113]\\
$\sigma$-Orionis & 34 $\pm$ 6 [71] & 14 $\pm$ 4 [74] & 34 $\pm$ 6 [71] & \ldots [0]\\
Serpens & 57 $\pm$ 4 [129] & 35 $\pm$ 4 [131] & 58 $\pm$ 5 [128] & 57 $\pm$ 7 [56]\\
Taurus & 61 $\pm$ 4 [202] & 39 $\pm$ 3 [214] & 62 $\pm$ 3 [197] & 76 $\pm$ 6 [49]\\
Upper Sco & 9 $\pm$ 2 [250] & 4 $\pm$ 1 [250] & 10 $\pm$ 2 [232] & 14 $\pm$ 3 [186]\\
\hline
AB Dor & 4 $\pm$ 2 [77] & 4 $\pm$ 2 [77] & 1 $\pm$ 2 [77] & 0 $\pm$ 1 [50]\\
Argus & 0 $\pm$ 2 [45] & 0 $\pm$ 1 [44] & 0 $\pm$ 2 [44] & \ldots [3]\\
$\beta$ Pic & 8 $\pm$ 4 [50] & 4 $\pm$ 3 [50] & 6 $\pm$ 3 [49] & 14 $\pm$ 5 [43]\\
Carina & 4 $\pm$ 3 [28] & 0 $\pm$ 1 [27] & 4 $\pm$ 3 [28] & \ldots [6]\\
Columba & 2 $\pm$ 2 [57] & 0.0 $\pm$ 0.9 [57] & 2 $\pm$ 2 [57] & 10 $\pm$ 8 [20]\\
$\epsilon$ Cha & 21 $\pm$ 8 [24] & 17 $\pm$ 7 [24] & 17 $\pm$ 8 [24] & 20 $\pm$ 10 [13]\\
Octantis & 0 $\pm$ 3 [17] & 0 $\pm$ 2 [17] & 0 $\pm$ 3 [17] & \ldots [1]\\
Tuc-Hor & 6 $\pm$ 3 [48] & 6 $\pm$ 3 [48] & 0 $\pm$ 1 [48] & 2 $\pm$ 3 [41]\\
TW Hya & 20 $\pm$ 9 [20] & 0 $\pm$ 3 [20] & 20 $\pm$ 9 [20] & 20 $\pm$ 10 [14]\\
\hline
\end{tabular}
\tablefoot{The number of sources used is indicated within brackets. Sub-indexes reference fractions at different wavelength ranges, as defined in R14}
\end{center}
\end{table*}

\begin{table*}
  \caption{Values and errors obtained from fitting and exponential decay to disk fractions with time.}
  \label{tab:pp_debris_previous}
  \begin{center}
    \begin{tabular}{l c c c }
      \hline\hline 
Wavelength range & A & $\tau$ & C  \\
& (\%/Myr) & (Myr) & (\%) \\ 
\hline
Short & 60 $\pm$ 10 & 2.7 $\pm$ 0.7 & 1.1 $\pm$ 0.9\\
Intermediate & 90 $\pm$ 10 & 3.3 $\pm$ 0.6 & 1 $\pm$ 1\\
Long (primordial) & 84 $\pm$ 6 & 4.4 $\pm$ 0.5 & -0.1 $\pm$ 0.4\\
Long & 95 $\pm$ 8 & 5.3 $\pm$ 0.9 & 0 $\pm$ 1\\
\hline
\end{tabular}
    \end{center}
\tablefoot{The long (primordial) case corresponds to red circles in Fig.~\ref{fig:pp_debris_previous}.}
\end{table*}

\begin{figure*}
\centering
\caption{Upper left panel: comparison of the disk fractions obtained in this study (circles) with those from  \citep{Hernandez2007,Hernandez2008} (squares) for regions in common.  Upper right and lower panels: disk fractions for all regions in the short, intermediate and long wavelength ranges. The best-fit exponential law is over-plotted (red line). }\label{fig:diskfractions_previous}
\includegraphics[width=0.7\textwidth]{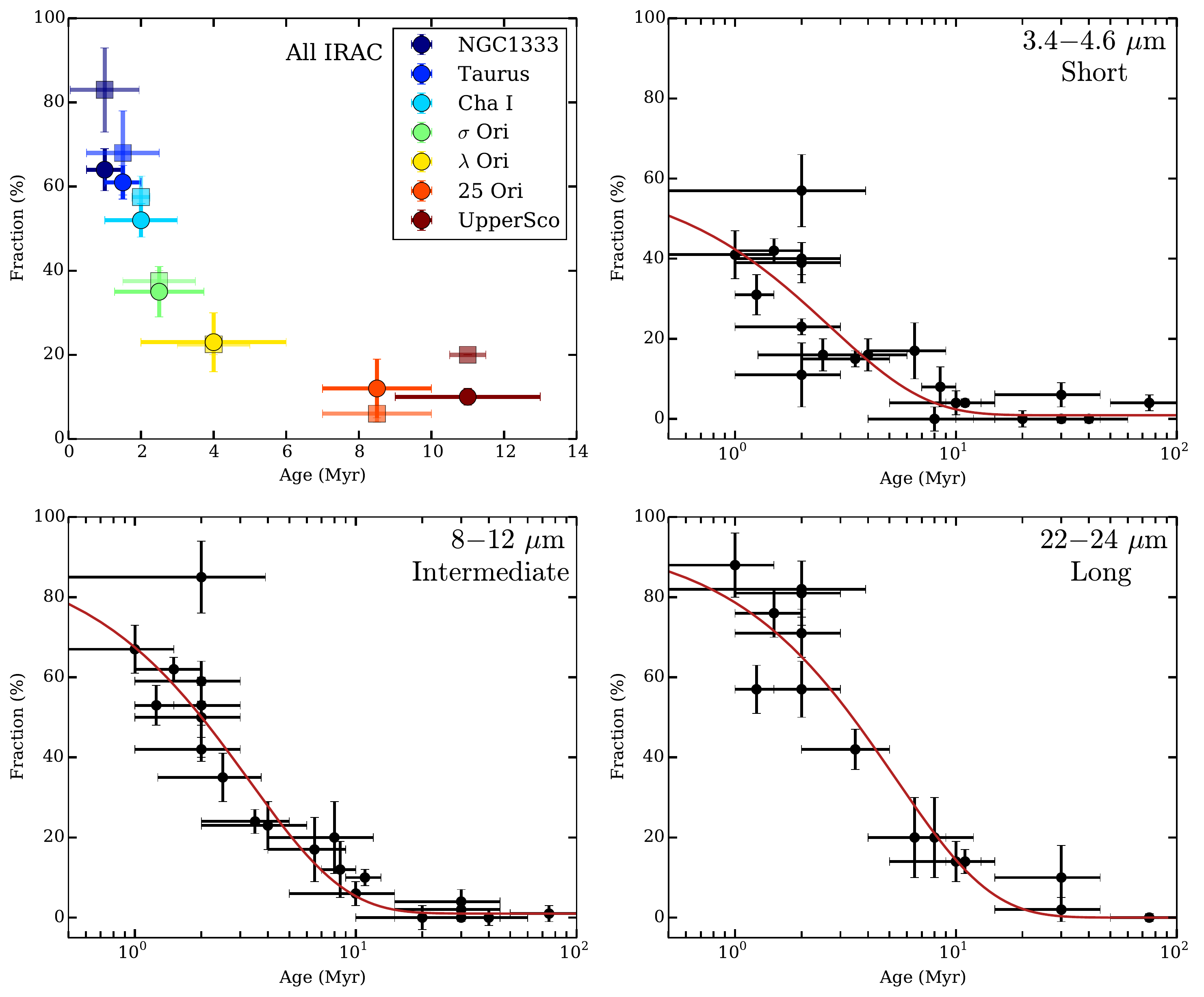}
\end{figure*}

\begin{figure*}
\centering
\caption{Fraction of primordial (red circles) and evolved (blue squares) disks as a function of age. The best-fit exponential law for the primordial disk percentages is also shown (red dashed line).}\label{fig:pp_debris_previous}
\includegraphics[width=0.5\textwidth]{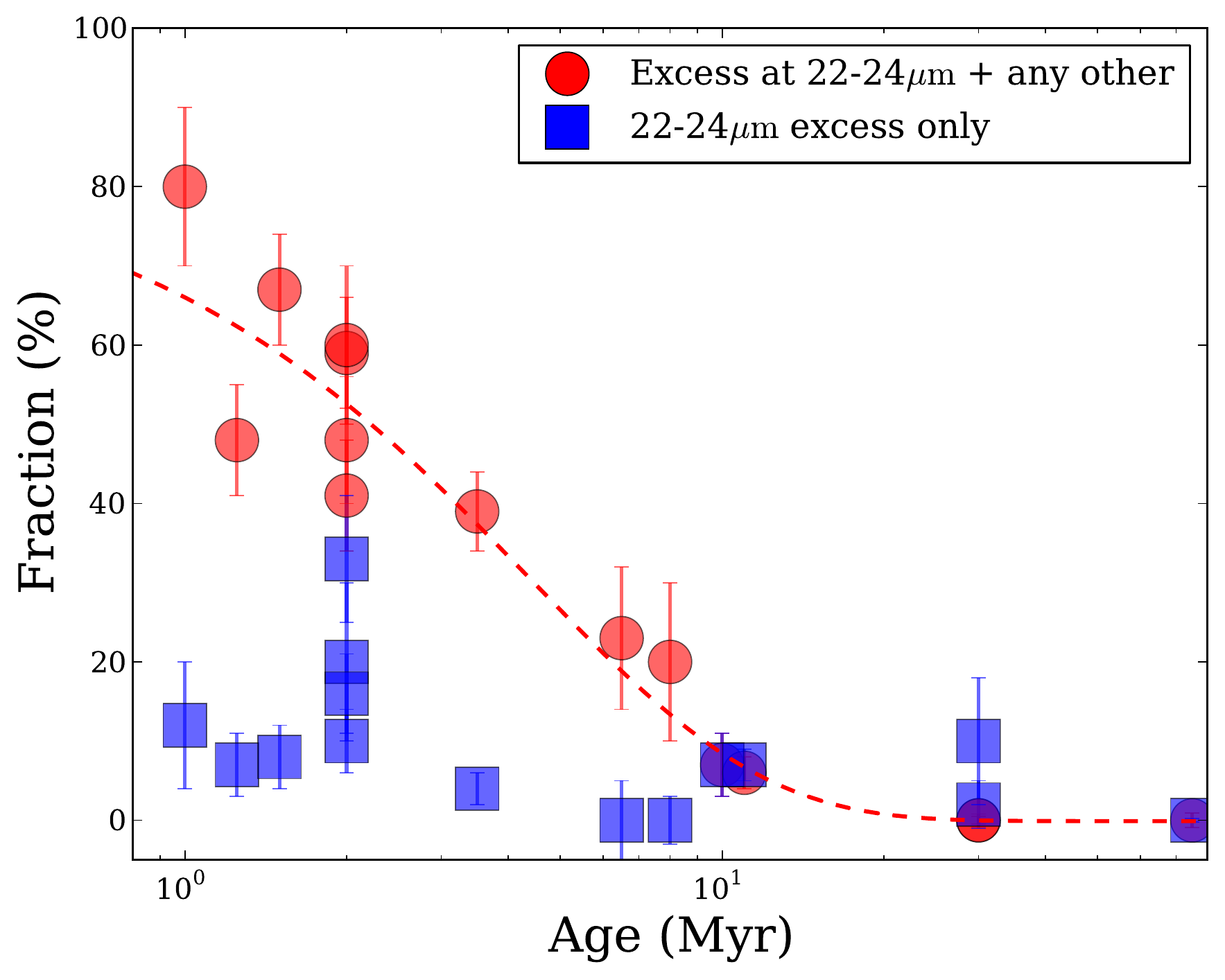} 
\end{figure*}

\end{appendix}

\end{document}